\documentclass[]{interact}
\usepackage[dvipsnames]{xcolor}
\usepackage{epstopdf}% To incorporate .eps illustrations using PDFLaTeX, etc.
\usepackage[caption=false]{subfig}% Support for small, `sub' figures and tables
\usepackage{braket, amsmath}
\usepackage[numbers,sort&compress]{natbib}% Citation support using natbib.sty
\bibpunct[, ]{[}{]}{,}{n}{,}{,}% Citation support using natbib.sty
\makeatletter% @ becomes a letter
\def\NAT@def@citea{\def\@citea{\NAT@separator}}% Suppress spaces between citations using natbib.sty
\makeatother% @ becomes a symbol again

\theoremstyle{plain}% Theorem-like structures provided by amsthm.sty

\theoremstyle{definition}

\theoremstyle{remark}

\begin{document}
	
	%\articletype{ARTICLE TEMPLATE}% Specify the article type or omit as appropriate
	
	\title{A Geometry of entanglement and entropy}
	
	\author{
		\name{Ramita Sarkar\textsuperscript{a,b}\thanks{CONTACT Ramita Sarkar. Email: ramitasarkar11@gmail.com}, Soumik Mahanti\textsuperscript{a,c} and Prasanta K. Panigrahi\textsuperscript{a}}
		\affil{\textsuperscript{a} Indian Institute of Science Education and Research Kolkata, Mohanpur 741246, West Bengal, India; \textsuperscript {b} Institute of Physics Bhubaneswar, Odisha, India - 751005; \textsuperscript {c} S.N. Bose National Center for Basic Sciences, Block JD, Sector III, Salt Lake, Kolkata 700106, India}
	}
	
	\maketitle
	
	\begin{abstract}
		This paper explores the fundamental relationship between the geometry of entanglement and von Neumann entropy, shedding light on the intricate nature of quantum correlations. We provide a comprehensive overview of entanglement, highlighting its crucial role in quantum mechanics. Our focus centers on the connection between entanglement, von Neumann entropy, a measure of the information content within quantum systems and the geometry of composite Hilbert spaces. We discuss various methods for quantifying and characterizing entanglement through a geometric perspective and elucidate how this connection unveils the nature of quantum entanglement, offering valuable insights into the underlying structure of quantum systems. This study underscores the significance of geometry as a key tool for understanding the rich landscape of quantum correlations and their implications across various domains of physics and information theory. An example of entanglement as an indispensable resource for the task of state teleportation is presented at the end.
	\end{abstract}
	

	\section{Introduction}
	
 In the standard formulation of Quantum theory, a state is a positive semi-definite linear hermitian operator acting on the Hilbert space having unit trace. In mathematical language, $\rho \in \mathcal{L(H)}$ satisfying $\rho=\rho^\dagger$, $\rho \geq 0$, and $Tr(\rho) = 1$ is a valid quantum state. As states are linear operators, any convex combination of states $\rho ^\prime = \sum_i p_i\rho_i$ will still be a valid state. The state that satisfies the property $\rho^2=\rho$ are termed as pure states. Pure states can be written down as projectors $\ket{\psi}\bra{\psi}$, where $\ket{\psi}$ represents the usual state vector of the Hilbert space, and its projection on the position Hilbert space $\braket{x|\psi}$ gives the familiar wavefunction $\psi(x)$ of the quantum mechanics. The Hilbert space of a composite system is combined by the tensor product of the individual Hilbert spaces according to the postulates of quantum mechanics. To illustrate, the full wavefunction of an electron inside an atom represents a vector in the Hilbert space $\mathcal{H}^n\otimes\mathcal{H}^l\otimes\mathcal{H}^s\otimes\mathcal{H}^{m_s}$, where $n$ denotes the principal quantum number representing the orbit, $l$, $s$, and $m_s$ being the orbital angular momentum, spin angular momentum, and spin quantum number respectively. Each individual space represents different degrees of freedom of a single electron, however, the complete wavefunction is an element of a composite Hilbert space. It is quite obvious that a state of a composite Hilbert space $\rho^{AB} \in \mathcal{L(H^A\otimes H^B)}$ cannot always be broken down as an operator on $\mathcal{L(H^A)\otimes L(H^B)}$. In particular, the states that can be expressed in the latter form are called product states, whereas the states that are a convex mixture of product states are called \emph{separable}. The states that are not separable, are known as \emph{entangled states}, and the property of non-separability across a composite Hilbert space is known as entanglement. In fact, as the local dimension for the individual Hilbert spaces or the number of spaces increases, much more often than not this breaking down will be impossible. This can be illustrated in the following way - consider an arbitrary pure quantum state of $n$ spin-$\frac{1}{2}$ systems $\ket{\phi}$. If the basis of the individual spin-$\frac{1}{2}$ systems are denoted as $\ket{0}$ and $\ket{1}$, the state can be represented as 

$$\ket{\phi} = \sum_{i_1,i_2,\dots, i_n \in \{0,1\}}c_{i_1i_2\dots i_n} \ket{i_1i_2\dots i_n}.$$  

\noindent The normalization condition translates to $\sum_{i_1,i_2,\dots, i_n \in \{0,1\}}|{c_{i_1i_2\dots i_n}|^2 =1}$. The number of complex variables scales as $2^n$ with the number of spin systems $n$. Analogously an arbitrary product state can be written down as 

$$\ket{\phi_{prod}} = (\alpha_1 \ket{0}+\beta_1\ket{1})\otimes (\alpha_2 \ket{0}+\beta_2\ket{1})\otimes \dots\otimes (\alpha_n \ket{0}+\beta_n\ket{1})$$

 Hence, the arbitrary product state of $n$ spin-$\frac{1}{2}$ systems have $2n$ complex variables. Therefore, as one increases the system size $n$, the size of the state space grows exponentially, whereas the space size of the product state grows only linearly; consequently, more and more of any randomly chosen state will be entangled state. Therefore, entanglement is not a rarity, but more of a norm. Entanglement has many peculiar consequences that are otherwise not seen in the classical macroscopic world. In 1935, Schr\"odinger noticed that~\cite{schrodinger19801}, while we have all the information about the joint state of a cat in a radioactive chamber represented as $\frac{1}{\sqrt{2}}(\ket{0; alive}+\ket{1; dead})$, no information about the individual constituent is known. This would imply that a cat is both dead and alive, which is an absurd conclusion. In mathematical terms, the reduced state for each constituent for such a state would be proportional to identity. Therefore, there is no extractable information from the subsystems. In terms of von Neumann entropy, the full state has zero entropy as it is a pure state, but the partial subsystems have the highest entropy for these states. In other words, the reduced states of the partial subsystem are maximally mixed. This is the reason that Schr\"odinger noticed for entangled states like this, the information lies in the correlation between the constituents, rather than with the individuals themselves. As explained through the previous examples, different degrees of freedom for the same particle can be entangled, like spin and path for a light quantum, or so can be spatially separated different particles, like an electron and a positron from a pair-production of a gamma photon. The former is called intra-partite entanglement, while the latter is known as inter-partite entanglement. It is mostly the latter that has found extensive applications in many areas of promising technological advancements. Entanglement is not only a mathematical structure of the Hilbert space, revealing a degree of non-separability, but also has a physical manifestation. As we note, the state of spin-$\frac{1}{2}$ particles like electrons can be combined so that a two-particle system can have spin angular momentum of 0 or 1. Following the conventional literature, each particle can be at \emph{up} state or \emph{down} state. In a two-particle system, the spin angular momentum of ground state $\frac{1}{\sqrt{2}}(\ket{01}-\ket{10})$ is spin 0. Two electrons in a hydrogen molecule and helium atom ground state are in this singlet state. The singlet state became the subject of much more investigation after the EPR argument~\cite{EPR1935} was generalized for spin-half systems by David Bohm~\cite{bohm2012quantum}. It is one of the well-known four Bell states, which are maximally entangled. Below are the explicit forms of the four- 
\begin{align*}
	\ket{\phi^+} &= \frac{1}{\sqrt{2}} (\ket{00}+\ket{11})\\
	\ket{\phi^-} &= \frac{1}{\sqrt{2}} (\ket{00}-\ket{11})\\
	\ket{\psi^+} &= \frac{1}{\sqrt{2}} (\ket{01}+\ket{10})\\
	\ket{\psi^-} &= \frac{1}{\sqrt{2}} (\ket{01}-\ket{10})
\end{align*}
Two of these maximally entangled Bell states manifest as singlet and triplet states of angular momentum, the former being a state immune to decoherence. From the point of view of quantum technologies, entangled states are an excellent resource for many purposes like Quantum communication~\cite{Bennet1992PRL,Bennet1993Teleport,gisin2007quantum}, Quantum Secret sharing~\cite{Hillery1999PRA}, Quantum Key Distribution~\cite{Gisin2002Rev_Mod_Phys}, Quantum computing~\cite{Raussendorf2001PRL,caleffi2022distributed} etc. The more entangled a state is, the more quantum advantage can be extracted in those protocols. Therefore, quantification of entanglement has a huge practical importance. Moreover, entanglement has a close connection with entropy and has many similarities with it~\cite{Popescu1997PRA,Horodecki2002PRL,Brandao2008Nature_Phys}. For example, entanglement is non-increasing under local operation and classical communication (LOCC). However, recently it has been shown that there is \emph{no} second law of entanglement manipulation~\cite{Lami_2023}, contrary to what was believed. In this article, we discuss the intricate connection between entanglement, entropy and geometry, which gives a deeper insight into the fundamental nature of entanglement as well as peeks into its distributive nature in a multipartite setting.

\section{Two qubit Entanglement} Any two-level quantum systems are called qubits, with the simplest example being the spin of an electron or the polarization of a photon. Conventionally, the two levels are represented as $\ket{0}$ and $\ket{1}$. They both form a basis in the $\mathcal{C}^2$ vector space, which is represented in the column vector form as $\begin{pmatrix}
	1\\
	0
\end{pmatrix}$ and 
$\begin{pmatrix}
	0\\
	1
\end{pmatrix}$ respectively. In this form, the Bell states read like the following
\begin{align*}
	\ket{\phi^\pm} &= \frac{1}{\sqrt{2}} \begin{pmatrix}
		1\\
		0\\
		0\\
		\pm 1
	\end{pmatrix}\\ 
	\ket{\psi^\pm} &= \frac{1}{\sqrt{2}} \begin{pmatrix}
		0\\
		1\\
		\pm 1\\
		0
	\end{pmatrix}\\ 
\end{align*}
If we write down the column vector in a matrix form with the first two entries as the element of the first column, and the next two entries as the second, then the matrix form for the Bell states becomes the following
$$ \ket{\phi^\pm}\rangle = \frac{1}{\sqrt{2}} \begin{pmatrix}
	1 & 0\\
	0 & \pm 1
\end{pmatrix}$$
$$ \ket{\psi^\pm}\rangle = \frac{1}{\sqrt{2}} \begin{pmatrix}
	0 & 1\\
	\pm 1 & 0
\end{pmatrix}$$
In both cases, the absolute value of the determinant of the matrices is $\frac{1}{2}$. Instead, if we had started with a product state like $\ket{00}$, $\ket{01}$ etc, evidently the matrix would have only one non-zero element and the determinant would have been 0. If the state is not maximally entangled but of the form $\frac{1}{\sqrt{3}}(\ket{00}+\ket{01}+\ket{10})$, the matrix would look like $\frac{1}{\sqrt{3}} \begin{pmatrix}
	1 & 1\\
	1 & 0
\end{pmatrix}$ The absolute value of its determinant would give $\frac{1}{3}$. This observation allows us to quantify entanglement for two-qubit systems which is illustrated in the following section.

\section{Entanglement Measure for two qubit states}
We consider a two-qubit system with qubits A and B. Let $\ket{\psi}$ be a normalized pure state of the system with
\begin{equation}\label{2e1}
	\ket{\psi}= a \ket{0_A0_B}+ b\ket{0_A1_B} +c \ket{1_A0_B}+ d \ket{1_A1_B}, 
\end{equation}
where $a,b,c,d \in \mathcal{C}$ satisfying the normalization condition.	We can rewrite \ref{2e1} as
\begin{align}\label{2e2}
	\ket{\psi}&= \ket{0_A}(a \ket{0_B}+ b\ket{1_B}) + \ket{1_A}(c \ket{0_B}+ d \ket{1_B}), \text{or}\\
	\ket{\psi}&=  \ket{0_A}\braket{0_A |\psi}+ \ket{1_A}\braket{1_A|\psi}.
\end{align} 
It is obvious from the algebra that the state can be expressed in a product form only if $\braket{0_A |\psi} = m\braket{1_A |\psi}$ for some complex number $m$. $\braket{0_A |\psi}$ and $\braket{1_A |\psi}$ are called un-normalized post-measurement vectors for subsystem $B$. In other words, the bipartition $A|B$ is separable when the vectors  $\braket{0_A |\psi}=a \ket{0_B}+ b\ket{1_B}$ and $\braket{1_A|\psi}=c \ket{0_B}+ d \ket{1_B} $ are parallel; precisely, $$\frac{a}{c} = \frac{b}{d}$$ or $ad-bc=0$ \cite{bhaskara2017generalized}.  Let, $\frac{a}{c} = \frac{b}{d} = k$, then $a =ck, \text{ and } b = dk$. In column vector form, the state is expressed as $\begin{pmatrix}
	a\\
	b\\
	c\\
	d\\
\end{pmatrix} = \begin{pmatrix}
	ck\\
	dk\\
	c\\
	d\\  
\end{pmatrix}$ = $\begin{pmatrix}
	k\\
	1
\end{pmatrix} \bigotimes \begin{pmatrix}
	c\\
	d
\end{pmatrix}$.  
Therefore, parallelism of post-measurement vectors in a subsystem implies separability across a bipartition and vice-versa. We show next that the degree of non-parallelism of vectors leads to a measure of non-separability, i.e., entanglement. The matrix form for the vector is $\begin{pmatrix}
	a & b\\
	c& d
\end{pmatrix}$, and the absolute value of the determinant is $|ad-bc|$. If this $2\times 2$ matrix is full-rank, then the state is entangled, otherwise, it is separable. In the most general bipartite scenario, a pure state in $\mathcal{C}^{d_1}\otimes \mathcal{C}^{d_2}$ can be expressed in a matrix form of the dimension $d_1\times d_2$. The state will be separable only if the matrix is of rank 1. In terms of post-measurement vectors, this implies all the vectors must be parallel. We show that the determinant of this matrix, $|ad-bc|$ in the $2 \times 2$ case, can be a faithful measure of entanglement. Also, we can express entanglement of a state as 2 times of the wedge product of the post measurement vectors, $ 2|\braket{0_A |\psi} \wedge \braket{1_A |\psi}| $  . The degree of non-parallelism of the vectors can be quantified through the wedge product between them~\cite{bhaskara2017generalized}. In $\mathcal{C}^2$, the wedge product between vectors is the generalization of cross-product to complex variables and it amounts to the area of the parallelogram formed by these two vectors. For achieving maximum entanglement, the parallelogram formed by the vectors $\braket{0_A |\psi}$ and $\braket{1_A|\psi}$ must have the maximum possible area, hence, it must be square. Keeping the normalisation condition in mind ($ |\braket{0_A |\psi}|^2+|\braket{1_A |\psi}|^2=1$), the maximum possible area can be found when the side of the square is equal to $\frac{1}{\sqrt{2}}$. \\
We can also verify that the parallelogram formed by the Bell states is a square. Bell states are written in the following form
\begin{equation}\label{bell}
	\ket{\phi^\pm}= \frac{\ket{00} \pm \ket{11}}{\sqrt{2}},  \\
	\ket{\psi^\pm}= \frac{\ket{01} \pm \ket{10}}{\sqrt{2}}
\end{equation}
Vectors formed by the Bell state $\ket{\phi^+}$ are $\braket{0_A |\phi^+}=\frac{1}{\sqrt{2}} \ket{0_B}$ and $\braket{1_A|\phi^+}=\frac{1}{\sqrt{2}} \ket{1_B} $. Thus the modulus of the each side is $\frac{1}{\sqrt{2}}$ and $\ket{0_B}$
and $\ket{1_B}$ are perpendicular, they will form a square.
\begin{figure}[h]
	\centering
	\includegraphics[scale=0.3]{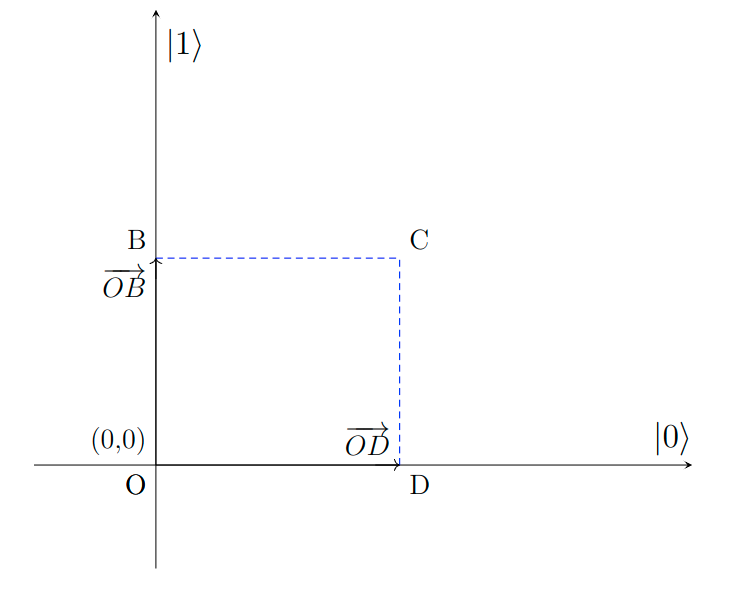}
	\caption{Area spanned by the vectors $\vec{OB}$ and $\vec{OD}$ of the Bell state $\ket{\phi^+}$ with equal sides and OBCD is a square. }
	\label{bellfig}
\end{figure}
Any two-party system can be written in a Schmidt decomposed form. This allows a single parameter to determine the measure of entanglement. Because of that, all two-party entanglement measures are a monotonic function of reduced subsystem's von Neumann entropy. However, as one goes to more than two-party case, there are more than one entanglement parameter, and thus there is no unique way to quantify entanglement.

If $\ket{\psi}$ be a pure state of a general $3$-qubit system with qubits A, B, C, then one can write, 
\begin{equation}\label{3e1}
\ket{\psi}= a \ket{000}+ b\ket{001} +c \ket{010}+ d \ket{011} + p \ket{100}+ q \ket{101}+ r \ket{110}+ s \ket{111}.
\end{equation}
We then consider the bi-partition A$\lvert$ BC and measure system BC in the computational basis. The post-measurement non-normalised vectors for system A will be,
$\chi_{0}^A= a\ket{0} + p\ket{1}, 
\chi_{1 }^A = b\ket{0} + q\ket{1},
\chi_{2}^A= c\ket{0} + r\ket{1}$, and  
$\chi_{3}^A= d\ket{0} + s\ket{1}$.
The squared concurrence in this bi-partition of the system is given by \cite{bhaskara2017generalized, mahanti2022classification, sarkar2022geometry, dutta2019permutation}, 

\begin{multline}\label{3e2}
C^{2}_{A\lvert BC}=4[\lvert \chi_{0}^A \wedge \chi_{1}^A \rvert^2 +\lvert \chi_{0}^A \wedge \chi_{2}^A \rvert^2+\lvert \chi_{0}^A \wedge \chi_{3}^A \rvert^2 +\lvert \chi_{1}^A \wedge \chi_{2}^A \rvert^2+\lvert \chi_{1}^A \wedge \chi_{3}^A \rvert^2+\lvert \chi_{2}^A \wedge \chi_{3}^A \rvert^2 ].
\end{multline}
Similarly, for the bi-partition B$\lvert$ CA, we measure system BC on a computational basis. The post-measurement non-normalised vectors for system B will be \cite{banerjee2020quantifying},
$\chi_{0}^B= a\ket{0} + c\ket{1}, 
\chi_{1 }^B = b\ket{0} + d\ket{1},
\chi_{2}^B= p\ket{0} + r\ket{1}$, and  
$\chi_{3}^B= q\ket{0} + s\ket{1}.$
The squared concurrence in this bi-partition of the system is given by, 

\begin{multline}\label{3e3}
C^{2}_{B\lvert CA}=4[\lvert \chi_{0}^B \wedge \chi_{1}^B \rvert^2 +\lvert \chi_{0}^B \wedge \chi_{2}^B \rvert^2+\lvert \chi_{0}^B \wedge \chi_{3}^B \rvert^2 +\lvert \chi_{1}^B \wedge \chi_{2}^B \rvert^2+\lvert \chi_{1}^B \wedge \chi_{3}^B \rvert^2+\lvert \chi_{2}^B \wedge \chi_{3}^B \rvert^2 ]
\end{multline}	
for the bi-partition C$\lvert$ BA and measure system BC on the computational basis. The post-measurement non-normalised vectors for system A will be,
$\chi_{0}^C= a\ket{0} + b\ket{1}, 
\chi_{1 }^C = c\ket{0} + d\ket{1},
\chi_{2}^C= p\ket{0} + q\ket{1}$, and  
$\chi_{3}^C= r\ket{0} + s\ket{1}$.
The squared concurrence in this bi-partition of the system is given by, 

\begin{multline}\label{3e4}
C^{2}_{C\lvert AB}=4[\lvert \chi_{0}^C \wedge \chi_{1}^C \rvert^2 +\lvert \chi_{0}^C \wedge \chi_{2}^C \rvert^2+\lvert \chi_{0}^C \wedge \chi_{3}^C \rvert^2  +\lvert \chi_{1}^C \wedge \chi_{2}^C \rvert^2+\lvert \chi_{1}^C \wedge \chi_{3}^C \rvert^2+\lvert \chi_{2}^C \wedge \chi_{3}^C \rvert^2 ].
\end{multline}	
Those three bipartite entanglements are not
completely independent \cite{qian2018entanglement}. In their work, the entanglement polygon inequality described that one entanglement cannot exceed the sum of the
other two: 
\begin{equation}\label{in1}
C_{A\lvert BC} \le C_{B\lvert CA}+C_{C\lvert AB}.
\end{equation}
Another stronger inequality was also derived \cite{PhysRevA.92.062345,PhysRevLett.127.040403}; 
\begin{equation}\label{in2}
C^{2}_{A\lvert BC} \le C^{2}_{B\lvert CA}+C^{2}_{C\lvert AB}
\end{equation}

Here, we are going to verify this inequality using geometry. First, we will verify this inequality for the general state. For the general state \ref{3e1}, we will verify the inequality of \ref{in2}. 
\begin{multline}
C^{2}_{B\lvert CA}+C^{2}_{C\lvert AB}-C^{2}_{A\lvert BC}= 4[2(ad-bc)^2+2(ps-qr)^2+(as+pd-br-qc)^2] \ge 0. 
\end{multline}
Thus, for the general 3 qubit state, we can verify the inequality.
We are considering an example for GHZ state, W state.
\section{GHZ state}
The mathematical form of the GHZ state is
\begin{equation}
\ket{GHZ}=\frac{\ket{000}+\ket{111}}{\sqrt{2}}.
\end{equation}
The post-measurement non-normalised vectors for system A for GHZ states are \\
$\chi_{0}^A= \frac{1}{\sqrt{2}}\ket{0}$, and  
$\chi_{3}^A= \frac{1}{\sqrt{2}}\ket{1}$.\\
The post-measurement non-normalised vectors for system B  are \\
$\chi_{0}^B= \frac{1}{\sqrt{2}}\ket{0}$, and  
$\chi_{3}^B= \frac{1}{\sqrt{2}}\ket{1}$.\\
The post-measurement non-normalised vectors for system C  are \\
$\chi_{0}^C= \frac{1}{\sqrt{2}}\ket{0}$, and  
$\chi_{3}^C= \frac{1}{\sqrt{2}}\ket{1}$.\\
Here, $\chi_{0}^A=\chi_{0}^B=\chi_{0}^C=\overrightarrow{OA}$ and $\chi_{1}^A=\chi_{1}^B=\chi_{1}^C=\overrightarrow{OB}$

\begin{figure}[h]
	\centering
	\includegraphics[scale=0.5]{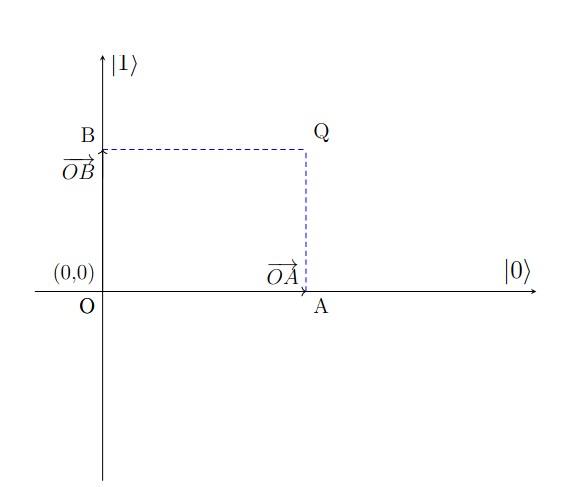}
	\caption{Area spanned by the vectors $\vec{OA}$ and $\vec{OB}$ }
	\label{1}
\end{figure}

The area spanned by the square OAQB  shown in Figure \ref{1} represents concurrence for the bipartition $A|BC$,
The same area will be obtained for the other two bipartitions $B|CA$ and $C|AB$. Now, we consider $C^{2}_{A\lvert BC}=4 \lvert \chi_{0}^A \wedge \chi_{3}^A \rvert^2=4 \lvert \vec{OA} \wedge \vec{OB} \rvert^2=M^2$, 
Then, immediately, we can check that \ref{in2} is verified. $ C^{2}_{B\lvert CA}+C^{2}_{C\lvert AB}=2M^2= 2C^{2}_{A\lvert BC}$.
In a similar fashion, \ref{in1} can be proved. 
$ C_{B\lvert CA}+C_{C\lvert AB}=2{M}= 2C_{A\lvert BC}$	
\section{W state}
\begin{figure}[]
	\centering
	\includegraphics[scale=0.5]{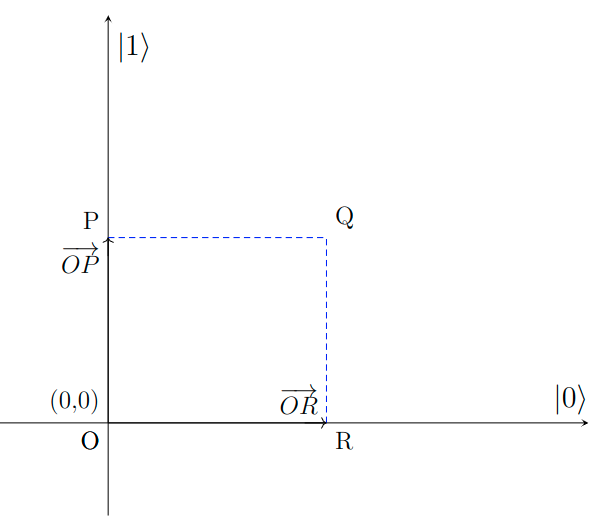}
	\caption{The area ORQP depicts the area formed by the vectors $\overrightarrow{OR}$ and $\overrightarrow{OP}$ i.e. $|\overrightarrow{OP}\wedge\overrightarrow{OR}|$ }
	\label{2}
\end{figure}

W state is defined as
\begin{equation}
\ket{W}=\frac{\ket{001}+\ket{010}+\ket{100}}{\sqrt{3}}.
\end{equation}
The post-measurement non-normalised vectors for system A for W states are \\
$\chi_{0}^A= \frac{1}{\sqrt{3}}\ket{1}$, 
$\chi_{1}^A= \frac{1}{\sqrt{3}}\ket{0}$, and  
$\chi_{2}^A= \frac{1}{\sqrt{3}}\ket{0}$.\\
The post-measurement non-normalised vectors for system B  are \\
$\chi_{0}^B= \frac{1}{\sqrt{3}}\ket{1}$,
$\chi_{1}^B= \frac{1}{\sqrt{3}}\ket{0}$, and  
$\chi_{2}^B= \frac{1}{\sqrt{3}}\ket{0}$.\\
The post-measurement non-normalised vectors for system C  are \\
$\chi_{0}^C= \frac{1}{\sqrt{3}}\ket{1}$, and  
$\chi_{1}^C= \frac{1}{\sqrt{3}}\ket{0}$, and  
$\chi_{2}^C= \frac{1}{\sqrt{3}}\ket{0}$.\\
Here, $\chi_{0}^A=\chi_{0}^B=\chi_{0}^C=\overrightarrow{OP}$ ;
$\chi_{1}^A=\chi_{2}^A=\chi_{1}^B=\chi_{2}^B=\chi_{1}^C=\chi_{2}^C=\overrightarrow{OR}$.	The area spanned for the bipartition $A|BC$ is shown in Figure \ref{2}. 
\\
Now, $C^{2}_{A\lvert BC}=4 [\lvert \chi_{0}^A \wedge \chi_{1}^A \rvert^2+\lvert \chi_{0}^A \wedge \chi_{2}^A \rvert^2]=2M^{\prime2}$ (say).\\
The same area will be obtained for the bipartition $B|CA$. \\
Here, $C^{2}_{B\lvert CA}=4 [\lvert \chi_{0}^B \wedge \chi_{1}^B \rvert^2+\lvert \chi_{0}^B \wedge \chi_{2}^B \rvert^2]=2M^{\prime2}$ .\\
For the bipartition, $C|AB$. $C^{2}_{C\lvert AB}=4 [\lvert \chi_{0}^C \wedge \chi_{1}^C \rvert^2+\lvert \chi_{0}^C \wedge \chi_{2}^C \rvert^2]=2M^{\prime2}$, 
Then, immediately, we can check that \ref{in2} is verified. $ C^{2}_{B\lvert CA}+C^{2}_{C\lvert AB}=4M^{\prime2}= 2C^{2}_{A\lvert BC}$.
Similarly, \ref{in1} can be proved. 
$ C_{B\lvert CA}+C_{C\lvert AB}=(2\sqrt{2})M^\prime= 2C_{A\lvert BC}$
\section{Qutrit state}
\begin{figure}[h!]
	\centering
	\includegraphics[width=0.50\linewidth]{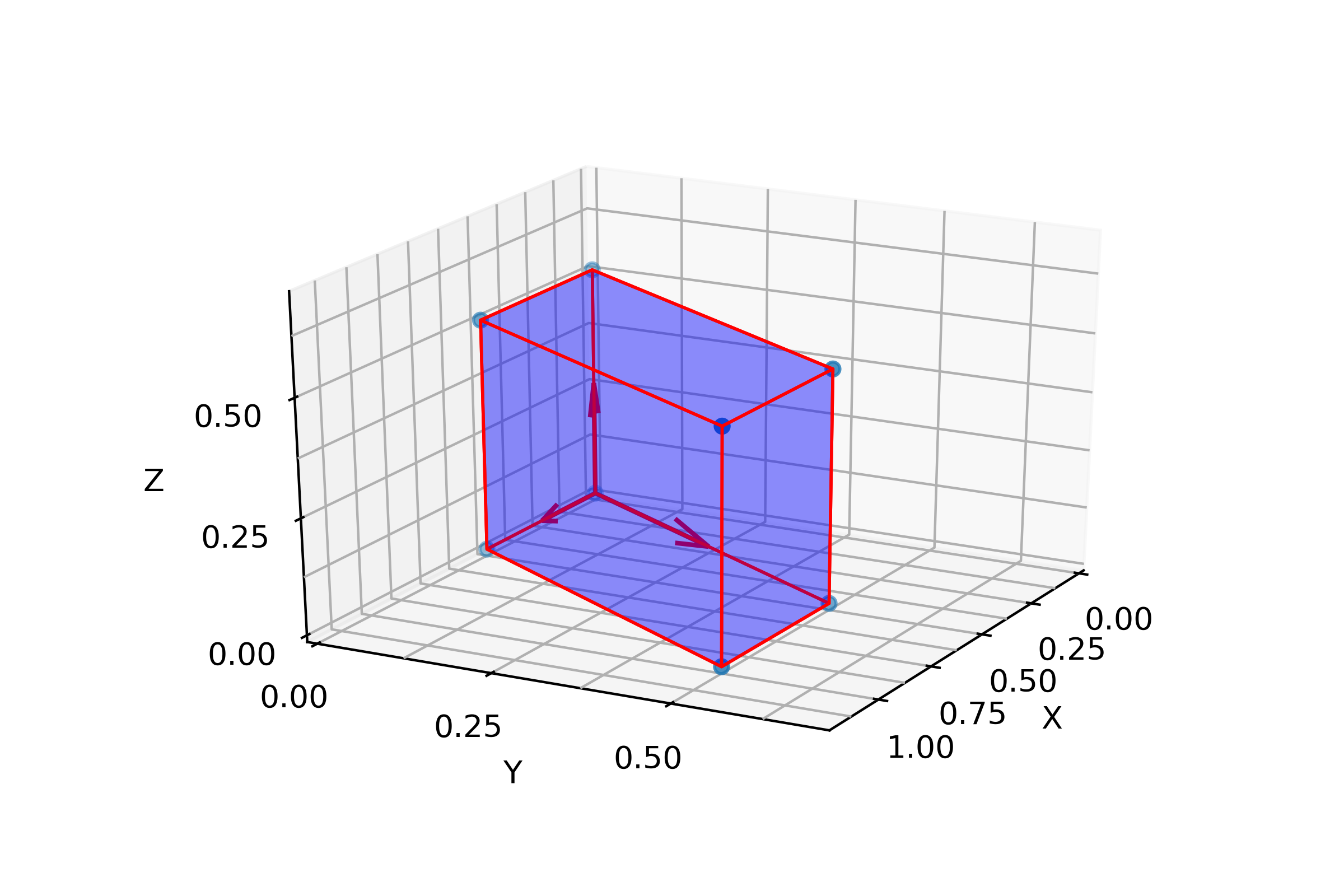}
	\caption{Total entanglement is volume plus the sum of all face areas}
	\label{fig:volume}
\end{figure}

A 3 qutrit general state can be written as,
\begin{multline}\label{q1}
\ket{\phi}=a_1\ket{000}+a_2\ket{001}+a_3\ket{002}+a_4\ket{010}+a_5\ket{011}+a_6\ket{012}+a_7\ket{020}+a_8\ket{021}+a_9\ket{022}\\
+a_{10}\ket{100}+a_{11}\ket{101}+a_{12}\ket{102}+a_{13}\ket{110}+a_{14}\ket{111}+a_{15}\ket{112}+a_{16}\ket{120}+a_{17}\ket{121}+a_{18}\ket{122}\\
+a_{19}\ket{200}+a_{20}\ket{201}+a_{21}\ket{202}+a_{22}\ket{210}+a_{23}\ket{211}+a_{24}\ket{212}+a_{25}\ket{220}+a_{26}\ket{221}+a_{27}\ket{222}.
\end{multline}
We consider the bi-partition A$\lvert$ BC and measure system BC on a computational basis. The post-measurement non-normalised vectors for system A will be,
$\chi_{0}^A= a_1\ket{0} +a_{10}\ket{1}+a_{19}\ket{2}, 
\chi_{1}^A= a_2\ket{0} +a_{11}\ket{1}+a_{20}\ket{2},
\chi_{2}^A= a_3\ket{0} +a_{12}\ket{1}+a_{21}\ket{2},  
\chi_{3}^A= a_4\ket{0} +a_{13}\ket{1}+a_{22}\ket{2},
\chi_{4}^A= a_5\ket{0} +a_{14}\ket{1}+a_{23}\ket{2},
\chi_{5}^A= a_6\ket{0} +a_{15}\ket{1}+a_{24}\ket{2},
\chi_{6}^A= a_7\ket{0} +a_{16}\ket{1}+a_{25}\ket{2},
\chi_{7}^A= a_8\ket{0} +a_{17}\ket{1}+a_{26}\ket{2}$, and
$\chi_{8}^A= a_9\ket{0} +a_{18}\ket{1}+a_{27}\ket{2}$.
The squared concurrence in this bi-partition of the system is given by 
\begin{multline}\label{q2}
C^{2}_{A\lvert BC}=4[\lvert \chi_{0}^A \wedge \chi_{1}^A \rvert^2 +\lvert \chi_{0}^A \wedge \chi_{2}^A \rvert^2+\lvert \chi_{0}^A \wedge \chi_{3}^A \rvert^2 +\lvert \chi_{0}^A \wedge \chi_{4}^A \rvert^2+\lvert \chi_{0}^A \wedge \chi_{5}^A \rvert^2+\lvert \chi_{0}^A \wedge \chi_{6}^A \rvert^2 +\lvert \chi_{0}^A \wedge \chi_{7}^A \rvert^2\\ +\lvert \chi_{0}^A \wedge \chi_{8}^A \rvert^2+\lvert \chi_{1}^A \wedge \chi_{2}^A \rvert^2+\lvert \chi_{1}^A \wedge \chi_{3}^A \rvert^2 +\lvert \chi_{1}^A \wedge \chi_{4}^A \rvert^2 +\lvert \chi_{1}^A \wedge \chi_{5}^A \rvert^2 +\lvert \chi_{1}^A \wedge \chi_{6}^A \rvert^2+\lvert \chi_{1}^A \wedge \chi_{7}^A \rvert^2\\+\lvert \chi_{1}^A \wedge \chi_{8}^A \rvert^2+\lvert \chi_{2}^A \wedge \chi_{3}^A \rvert^2+\lvert \chi_{2}^A \wedge \chi_{4}^A \rvert^2
+\lvert \chi_{2}^A \wedge \chi_{5}^A \rvert^2 +\lvert \chi_{2}^A \wedge \chi_{6}^A \rvert^2+\lvert \chi_{2}^A \wedge \chi_{7}^A \rvert^2+\lvert \chi_{2}^A \wedge \chi_{8}^A \rvert^2\\+\lvert \chi_{3}^A \wedge \chi_{4}^A \rvert^2+\lvert \chi_{3}^A \wedge \chi_{5}^A \rvert^2+\lvert \chi_{3}^A \wedge \chi_{6}^A \rvert^2+\lvert \chi_{3}^A \wedge \chi_{7}^A \rvert^2+\lvert \chi_{3}^A \wedge \chi_{8}^A \rvert^2
+\lvert \chi_{4}^A \wedge \chi_{5}^A \rvert^2+\lvert \chi_{4}^A \wedge \chi_{6}^A \rvert^2\\+\lvert \chi_{4}^A \wedge \chi_{7}^A \rvert^2+\lvert \chi_{4}^A \wedge \chi_{8}^A \rvert^2+\lvert \chi_{5}^A \wedge \chi_{6}^A \rvert^2+\lvert \chi_{5}^A \wedge \chi_{7}^A \rvert^2+\lvert \chi_{5}^A \wedge \chi_{8}^A \rvert^2+\lvert \chi_{6}^A \wedge \chi_{7}^A \rvert^2+\lvert \chi_{6}^A \wedge \chi_{8}^A \rvert^2+\lvert \chi_{7}^A \wedge \chi_{8}^A \rvert^2].
\end{multline}
Similarly, for the bi-partition B$\lvert$ CA and C$\lvert$ AB, we measure systems CA and AB on a computational basis. We will get post-measurement vectors in a similar fashion. In a recent work~\cite{mahanti2022classification}, the authors have used multiple wedge products between post-measurement vectors to construct an improved measure of entanglement which generalises the area of a parallelogram to the hypervolume of a parallelepiped.  Below is an example of the geometry of a parallelepiped constructed for the two-qutrit states in Figure~\ref{fig:volume}.

Below, we show examples of the geometry considering the special states, like qutrit GHZ state, W state, etc.
\section{Qutrit GHZ state}
The general form of the maximally entangled qutrit GHZ state is
\begin{equation}
\ket{GHZ}_{Qutrit}=\frac{\ket{000}+\ket{111}+\ket{222}}{\sqrt{3}}
\end{equation}
The post-measurement non-normalised vectors for system A for GHZ states are \\
$\chi_{0}^A= \frac{1}{\sqrt{3}}\ket{0}$,  
$\chi_{5}^A= \frac{1}{\sqrt{3}}\ket{1}$ and $\chi_{8}^A= \frac{1}{\sqrt{3}}\ket{2}$.\\
The post-measurement non-normalised vectors for system B  are \\
$\chi_{0}^B= \frac{1}{\sqrt{3}}\ket{0}$,  
$\chi_{5}^B= \frac{1}{\sqrt{3}}\ket{1}$ and $\chi_{8}^B= \frac{1}{\sqrt{3}}\ket{2}$.\\
The post-measurement non-normalised vectors for system C  are \\
$\chi_{0}^C= \frac{1}{\sqrt{3}}\ket{0}$,  
$\chi_{5}^C= \frac{1}{\sqrt{3}}\ket{1}$ and $\chi_{8}^C= \frac{1}{\sqrt{3}}\ket{2}$.\\
Here, $\chi_{0}^A=\chi_{0}^B=\chi_{0}^C=\overrightarrow{OA}$, $\chi_{5}^A=\chi_{5}^B=\chi_{5}^C=\overrightarrow{OB}$ 
and $\chi_{8}^A=\chi_{8}^B=\chi_{8}^C=\overrightarrow{OC}$

The area spanned by the square OAQB  shown in Figure \ref{1} represents concurrence for the bipartition,

The same area will be obtained for the three bipartitions $A|BC$, $B|CA$ and $C|AB$. Now, we consider $C^{2}_{A\lvert BC}=4 [\lvert \chi_{0}^A \wedge \chi_{5}^A \rvert^2+\lvert \chi_{0}^A \wedge \chi_{8}^A \rvert^2+\lvert \chi_{5}^A \wedge \chi_{8}^A \rvert^2]=4 [\lvert \vec{OA} \wedge \vec{OB} \rvert^2+\lvert \vec{OA} \wedge \vec{OC} \rvert^2+\lvert \vec{OB} \wedge \vec{OC} \rvert^2]=\frac{1}{3}=N^2$ (say), 
Then, immediately, we can check that \ref{in2} is verified. $ C^{2}_{B\lvert CA}+C^{2}_{C\lvert AB}=2N^2= 2C^{2}_{A\lvert BC}$.
In a similar fashion, \ref{in1} can be proved. 
$ C_{B\lvert CA}+C_{C\lvert AB}=2{N}= 2C_{A\lvert BC}$
\section{Entropy measure of entanglement for 2 and 3 qubit system}
The von Neumann entropy amounts to the mixedness of a state. If, for a pure state, the reduced subsystems are mixed, it indicates that the state is entangled. The more mixed the subsystems are, the more entangled the state will be. The entropy of entanglement or entanglement entropy is nothing but the von Neumann entropy of the reduced density matrix for any of the subsystems.
The von Neumann entropy, named after John von Neumann, is a generalisation of Shanon entropy. The von Neumann entropy for a quantum system with density matrix $\rho$ is defined as, 
\begin{equation}
S(\rho)= -Tr(\rho \ln \rho).
\end{equation}
The von Neumann entropy gives the realisation about the information present in a system as well as correlations among the sub-systems of the composite system.
We consider a two-qubit Bell state, for example. All of the four Bell states have zero entropy, while bipartitions of the states have entropy 1, maximally mixed. \\
Similarly, if we consider 3 qubit states, all of the pure states have zero entropy, but entangled states have non-zero entropy for the sub-systems. GHZ and W state both have zero entropy. Each sub-system of GHZ state has entropy 1, maximally mixed; Bipartitions of W state is 0.91. The reduced density matrix of the W state is not maximally mixed. We can study a variety of states in this regard. For multipartite states, the entropy of the subsystems follows several types of inequalities like the triangle inequality~\cite{Zhang2008PRA} or Araki-Lieb inequality~\cite{MORELLI2020294}. This limits how entangled one particle can be with another particle if it is already entangled to a third. These entropic inequalities give rise to the famous entanglement monogamy~\cite{Coffman2000PRA}. The constraints on the amount of entanglement between different parts of a system can be used to construct genuine multipartite entanglement measures, as shown in~\cite{Mahanti2023Geometric}.
\section{Teleportation: An Application of Entanglement}

Quantum teleportation can be achieved with perfect fidelity, owing to maximally entangled states being used as a resource. Otherwise, Bob achieves this probabilistically. An unknown state $\psi$, received by Alice, can be reproduced at Bob's end and is located at a distant point through the Bell state $\frac{\ket{00}+\ket{11}}{\sqrt{2}}$. This is done through the following procedure involving joint measurement of the Bell-type with one bit of classical communication from Alice, succeeded by local unitary operation at Bob's end. This procedure is called local operation and classical communication (LOCC), as Alice performs an operation in her labs only and uses a classical channel like a telephone call to send the classical information of the measurement outcome. More explicitly the composite state of the qubit to be sent and the shared entangled state is the following, i.e.,  $\ket{\psi}\otimes \ket{\phi^+} = (\alpha\ket{0}+\beta\ket{1})\otimes\frac{\ket{00}+\ket{11}}{\sqrt{2}}$. After the Bell measurement, the arbitrary outcome is one of the four terms in the above equation.

The state has been reproduced at Bob's end. It is to be noted that in the process, Alice has completely decoupled from Bob and fully entangled with the unknown state particle.  This reflects a peculiar property of entanglement known as monogamy. It is to be noted that the unknown state has not been sent to Bob's side but has been created at his side through the exact reproduction of the parameters $\alpha$ and $\beta$, in general, complex. One more point to note is that the outcome of Bell's measurement collapses the full wave function to one of the four possible eigenvectors (Bell states). One has to pass on their information to Bob. That is done through a conventional telephone line with the signal speed not exceeding the speed of light. Once Alice's measurement outcome is conveyed to Bob, who applies one of the four possible unitary operations, namely $I, \sigma_x, i\sigma_y, \text{ or } \sigma_z$ and gets back the desired state. It is although entanglement is overcoming the distance, information transfer can not take place faster than the speed of light. 
\section{Conclusion}
In this article, we describe the relationship between the geometry of entanglement and von Neumann entropy. We relate how geometry plays a vital role in entanglement and other quantum correlations like von Neumann entropy. We briefly define entanglement and its significance in quantum mechanics. We discuss the underlying geometry of entanglement starting from 2 qubit states, such as the Bell state; we end up with the geometry of 3 qutrit state. We describe various three-qubit pure states and their geometry, where we motivate that the area of the parallelogram of the post-measurement vectors of the states can be equivalent to the measure of entanglement.  We shed an idea of improved measurement of entanglement, which generalises the area of a parallelogram to the hypervolume of a parallelepiped. We give a detailed description of the von Neumann entropy of a multipartite pure state. An entangled pure state has zero von Neumann entropy, though its bipartition may have non-zero entropy. Finally, we give teleportation,  as an example of the application of entanglement. 
\section{Acknowledgement}  SM acknowledges support from the CSIR project 09/0921(16634)/2023-EMR-I. RS acknowledges PKP and IISER Kolkata for their hospitality during the academic visit. RS and SM acknowledge Rajiuddin Sk for his valuable discussion and suggestion.

\bibliographystyle{unsrt}	
\bibliography{lib}

%	\begin{thebibliography}{10}

%		\bibitem{iwata2019conditional}
%		Y.~Iwata and P.~Stevenson.
%		\newblock Conditional recovery of time-reversal symmetry in many nucleus
%		systems.
%		\newblock {\em New Journal of Physics}, (2019), vol. 21, 043010.
%		
	
%	\end{thebibliography}
\end{document}